\documentclass[11pt]{article}
\usepackage{xspace,amssymb,amsmath,helvet,graphicx}
\begin{document}
% useful commands
\newcommand{\dee}{\,\mbox{d}}
\newcommand{\naive}{na\"{\i}ve }
\newcommand{\eg}{e.g.\xspace}
\newcommand{\ie}{i.e.\xspace}
\newcommand{\pdf}{pdf.\xspace}
\newcommand{\etc}{etc.\@\xspace}
\newcommand{\PhD}{Ph.D.\xspace}
\newcommand{\MSc}{M.Sc.\xspace}
\newcommand{\BA}{B.A.\xspace}
\newcommand{\MA}{M.A.\xspace}
\newcommand{\role}{r\^{o}le}
\newcommand{\signoff}{\hspace*{\fill} Rose Baker \today}
% entry environment
\newenvironment{entry}[1]%
{\begin{list}{}{\renewcommand{\makelabel}[1]{\textsf{##1:}\hfil}%
\settowidth{\labelwidth}{\textsf{#1:}}%
\setlength{\leftmargin}{\labelwidth}
\addtolength{\leftmargin}{\labelsep}
\setlength{\itemindent}{0pt}
}}%
{\end{list}}
\title{Estimating accurate covariance matrices on fitted model parameters}
\author{Rose Baker\\School of Business\\University of Salford, UK}
%\email{rose.baker@cantab.net}
\maketitle
\begin{abstract}
The accurate computation of the covariance matrix of fitted model parameters is a somewhat neglected task in Statistics.
Algorithms are given for computing accurate covariance matrices derived from computing the Hessian matrix by numerical differentiation, and also
for the covariance matrix of the posterior distribution of model parameters. Evaluations on two datasets where the Hessian could be computed analytically
show that the numerical differentiation algorithm is very accurate.
\end{abstract}
\section*{Keywords}
Hessian; covariance matrix; maximum-likelihood; posterior density; adaptive importance sampling.
\section{Introduction}
Estimating model parameters is arguably the most important task in Statistics. If so, estimating the standard errors on fitted model parameters
would be the second most important task. Here we can distinguish two cases, the covariance matrix derived from the Hessian, and the covariance matrix for the 
posterior distribution of the $n$ model parameters $\boldsymbol\theta$.
In the first case, asymptotic arguments show that the desired covariance matrix ${\bf V}={\bf H}^{-1}$,
where the Hessian matrix
\[H_{ij}=-\partial^2 \ell/\partial\theta_i\partial\theta_j,\]
is the matrix of second derivatives of minus the log-likelihood $-\ell$ evaluated at the likelihood maximum. In the second case, we have Bayesian inference, in which the mean and covariance matrix of the posterior distribution of the
model parameters is desired. 

This article presents a method for accurately computing the covariance matrix derived from the Hessian, and also a method for computing expected-likelihood
means and covariance matrices. This work relies heavily on first maximising the likelihood and computing ${\bf V}$. Next, adaptive importance sampling (AIS)
with antithetic variation and control variates is used to estimate the posterior mean ${\tilde{\boldsymbol \theta}}$ and covariance matrix $\tilde{\bf V}$.
The methods have been prototyped in Fortran.

After fitting a model, \eg by maximum likelihood estimation (MLE), little attention has been paid to accurately estimating the covariance matrix for the fitted model parameters.
For standard problems, it is worthwhile for the software package writer to compute first and second derivatives analytically, which yields the Hessian. 
This removes the need for numerical computation of the Hessian. However, for non-standard problems
coding up first and second derivatives would be a burden, requiring often messy mathematics and programming. 
For example, if the likelihood function contains integrals, differentiation under the integral sign and the evaluation of further integrals will be required.
It is here that methods using only function evaluation are needed.

The R package nmle has a routine fdhess that evaluates the Hessian, approximating differentials as differences, and using the Koshal design to choose where to evaluate the function.
In this design, the function is evaluated with variables shifted singly by $\pm h$ for a step-length $h$, and with pairs of variables shifted by $h$.
It therefore requires few function evaluations for the off-diagonal Hessian elements.

It has been pointed out (Gower and Richt\'{a}rik, 2017) that the inverse Hessian can be approximately computed and updated during function minimization.
However, we are here concerned with computing accurate covariance matrices. Huang, Farewell and Pan (2017) give a method for patching up estimated covariance matrices that are not positive definite.
Here, we attempt to compute them accurately in the first place, and any patching (\eg eigenvalue polishing) uses additional function values.

Collinearity is one reason why the Hessian may not be invertible. Here we assume that this is not a problem, \ie the modeller has devised a sensible model.
However, the methods described here will produce a positive-definite covariance matrix even in poorly-determined situations.

\section{Covariance matrices for maximum-likelihood estimates}
There is no standard method for computing the covariance matrix, but the following method was developed, found to be accurate, and does not require too many function evaluations.
Accuracy was the first consideration, with controlling the number of function evaluations the second priority.
It is assumed that a function minimiser has found the maximum of the log-likelihood $\ell$, and that a function can return values of $f({\boldsymbol\theta})=-\ell({\boldsymbol\theta})$
at any chosen value of $\boldsymbol\theta$. 
The strategy was to compute accurate values of the $n$ second derivatives $\partial^2 f/\partial\theta_i^2$, which also yields the
optimum steplength $h_i$. For the $n(n-1)/2$ different off-diagonal derivatives, such accuracy is impossible without very many function evaluations.
Therefore, the optimum steplength was used, and the estimated off-diagonal Hessian element improved by applying Richardson extrapolation once.

The diagonal derivatives (curvatures) are computed using Richardson extrapolation via Ridders' method.
The central difference estimate
\[\dee^2 f(x)/\dee x^2|_{x=x_0} \simeq \frac{f(x_0+h)-2f(x_0)+f(x_0-h)}{h^2}\equiv c_h\]
has error $O(h^2)$, so that
\[c_h=\dee^2 f(x)/\dee x^2|_{x=x_0}+Ch^2+Dh^4+\cdots.\]
Halving $h$ (for example) and solving, we have that $\frac{4c_{h/2}-c_h}{3}$ has error $O(h^4)$ and that $Ch^2=(4/3)(c_h-c_{h/2})$.
This forms the basis of Ridders' method, in which $h$ is progressively scaled down by a factor of $\beta \simeq \sqrt{2}$ until the estimated error from rounding
starts to increase (\eg Press {\em et al} (2007)).

The full algorithm used is then:
\begin{enumerate}
\item If $\epsilon$ is machine accuracy, start with a small step $h=\epsilon^{1/4}$ and keep doubling until
$f(x_0+h) > f(x_0)$ and $f(x_0-h) > f(x_0)$.
\item estimate the curvature as $\sigma=h/\sqrt{f(c_0+h)-2f(x_0)+f(x_0-h)}$.
\item Apply Ridders' method starting with a large stepsize, \eg $h=\sigma/2$.
This delivers the $n$ curvatures $c_i$ and their optimum steplengths $h_i$.
\item To find the off-diagonal elements $H_{ij}$, a Richardson extrapolation is applied just once,
starting with stepsizes $h_i, h_j$ using the central difference approximation $D(h_x, h_y)$ of
\[\frac{f(x_0+h_x,y_0+h_y)+f(x_0-h_x,y_0-h_y)-f(x_0+h_x,y_0-h_y)-f(x_0-h_x,y_0+h_y)}{4h_x h_y}.\]
This has error $O(h_x h_y)$ and so after Richardson extrapolation we obtain the formula
\[ H_{ij} \simeq \frac{4D(h_x/2, h_y/2)- D(h_x, h_y)}{3}\]
which has error $O(h_x^2h_y^2)$.
\end{enumerate}

The rationale is that the $n$ diagonal elements can be estimated accurately, but we cannot be as prodigal with function evaluations for the off-diagonal elements.
Each off-diagonal matrix element is here given 8 function evaluations, 4 required for the central-difference approximation to $H_{ij}$, and 4 from the same approximation at half the step-lengths $h_x, h_y$.

Subsequently, the Hessian ${\bf H}$ must be inverted to give the covariance matrix ${\bf V}$. To ensure accuracy and that ${\bf V}$ is positive definite,
two methods are available: eigenvalue polish and recomputation in the diagonal frame. The eigenvalues ${\boldsymbol \xi}$ (elements of the diagonal matrix ${\bf D}$) and eigenvector matrix $\boldsymbol \Lambda$ of the Hessian are found, so that
${\bf H}={\boldsymbol \Lambda}{\bf D}{\boldsymbol \Lambda}^T$. The eigenvalues can be `polished' by recomputing them using the algorithm previously described.
It is also possible to recompute ${\bf D}$ entirely, although this is an expensive strategy. Eigenvalue polish ensures that ${\bf H}$ is positive definite and requires only $O(n)$
function evaluations.

The computation of the Hessian takes $O(n^2)$ operations, and its inversion $O(n^3)$ operations. However, this is very fast compared to the Hessian computation, so except for huge datasets, the methodology takes effectively $O(n^2)$ operations.

\section{Covariance matrix and mean for the posterior distribution of model parameters}
Bayesians want the mean and variance of the posterior distribution for $\boldsymbol\theta$ rather than the maximum-likelihood estimates.
It is also possible to regard such estimates in a frequentist sense as expected likelihood estimates, if vague prior distributions are used so that no subjective prior information is being included.
The complete class theorem indicates that such Bayesian-like estimators will be good frequentist estimators.

The mean will differ from the maximum likelihood estimate, and in general the variance of parameter estimates tends to be larger than the inverse Hessian.
Asymptotically, as sample size $N$ tends to infinity, the means and covariance matrices become identical. The expected likelihood estimates may be more appropriate than maximum likelihood estimates for small samples. 

For example, on estimating the mean and variance from a sample of $m$ independent and identically distributed observations from a normal distribution,
the maximum likelihood estimators would be $\bar{x}$ and $s^2=\sum_{i=1}^m (x_i-\bar{x})^2/m$ respectively. The expected likelihood estimates differ in that
the formula for the estimated variance is now $\hat{\sigma}^2=\sum_{i=1}^m (x_i-\bar{x})^2/(m-1)$, an unbiased estimator.

The only realistic way to evaluate the required $n$-dimensional integral when $n$ is large is by Monte-Carlo integration.
Press {\em et al} (2007) and Hesterberg (1996) discuss relevant techniques of variance reduction here.

Applied very naively, Monte-Carlo integration would give a formula for the mean
where ${\cal L}$ is the likelihood, and the ${\bf X}_i$ would be random numbers generated uniformly over a large hypercube.
The likelihood need only be known to a scale factor.

Using importance sampling improves the accuracy. Here we sample not uniformly,
but according to the `importance' of points ${\bf X}_i$ in evaluating the integral. The integrand is weighted to correct for this sampling scheme.

For approximate multivariate normality of the integrand, a multivariate normal sampling distribution could be used.
However, if the sampling distribution is shorter tailed than the likelihood function, very large weights can arise and make the integration inefficient.
it is therefore better to use a long-tailed distribution, and in this work, we take ${\bf Y}={\bf X}-\hat{\boldsymbol\theta}$ from a multivariate t distribution with $\nu$ degrees of freedom.
The weight for each observation then changes from $\cal L$ to $w_i={\cal L}(\hat{\boldsymbol\theta}+{\bf Y}_i)(1+\frac{{\bf Y}_i^Y{\bf H}{\bf Y}_i}{\nu-2})^{(n+\nu)/2}$
where $\nu > 2$ and the estimator is
\[\tilde{\boldsymbol\theta} \simeq \hat{\boldsymbol\theta}+\frac{\sum_{i=1}^N w_i Y_i}{\sum_{i=1}^N w_i}.\]
Three tricks are now used to speed up the computations.

First, antithetic variation is used whereby the likelihood is sampled at ${\bf Y}_i+\hat{\boldsymbol \theta}$ and $-{\bf Y}_i+\hat{\boldsymbol \theta}$.
Denote the weights by $w_{1i}$ and $w_{2i}$ respectively. Then
\[\tilde{\boldsymbol\theta} \simeq \hat{\boldsymbol\theta}+\frac{\sum_{i=1}^N (w_{1i}-w_{2i}) Y_i}{\sum_{i=1}^N (w_{1i}+w_{2i})}.\]
Clearly, if the log-likelihood were exactly normal $w_{1i}=w_{2i}$ and the correct answer of $\tilde{\boldsymbol\theta}=\hat{\boldsymbol \theta}$ would be obtained
after only one simulation.

To compute the covariance matrix $\tilde{\bf V}$, antithetic variation does not help as the quadratic terms do not change sign as ${\bf Y} \rightarrow -{\bf Y}$.
Instead, the method of control variates is used, the second trick.  The resulting formula is:
\begin{equation}\label{eq:vtilde}
\begin{split}
\tilde{\bf V} &\simeq \frac{(\nu-2)}{\nu}\frac{\sum_{i=1}^N (w_{1i}+w_{2i}){\bf Y}_i{\bf Y}_i^T}{\sum_{i=1}^N(w_{1i}+w_{2i})}\\
 &+\sum_{i=1}^N\frac{{\bf V}-\frac{(\nu-2)}{\nu}{\bf Y}_i{\bf Y}_i^T}{N}-(\tilde{\boldsymbol\theta}-\hat{\boldsymbol\theta})(\tilde{\boldsymbol\theta}-\hat{\boldsymbol\theta})^T
\end{split}
\end{equation}

In (\ref{eq:vtilde}) the third term corrects for the second moment being computed about $\hat{\boldsymbol\theta}$ instead of $\tilde{\boldsymbol\theta}$, which is necessary as both computations are done together term by term.
In the first term, the factor $\frac{(\nu-2)}{\nu}$ is needed to correct the observed second moment of the t-distribution to the parameter ${\bf H}^{-1}$.
The second term is the control term, with zero mean, which is negatively correlated with the first term. 

The third trick is to make the importance sampling adaptive,
by starting with a small value of $\nu$, say 4, performing $N$ simulations and computing the error, and then performing $N$ simulations with $\nu\rightarrow \sqrt{2}\nu$,
and so on. In practice, $\nu$ was rounded to the nearest integer. When the error increases, we revert to the previous value of $\nu$ for the remaining simulations.
Finally, all the $s$ (independent) estimates of a quantity $T$ are combined using the formula
\[T = \frac{\sum_{i=1}^s T_i/\sigma_i^2}{\sum_{i=1}^s 1/\sigma_i^2},\]
where the individual estimates $T_i$ have variance $\sigma_i^2$.

As we are evaluating many integrals at once, which error should we seek to minimise? The error on the trace of the covariance matrix was chosen.

Generating random numbers from a multivariate t-distribution can be done as follows: a Cholesky decomposition is done on the covariance matrix ${\bf V}$,
so that ${\bf V}={\bf L}{\bf L}^T$, where ${\bf L}$ is a lower-triangular matrix. Then if ${\bf X}$ is a vector of standard normal random variates, 
${\bf Y}={\bf L}{\bf X}$ is multivariate normal with covariance matrix ${\bf V}$. Finally, if $\nu$ is integer, ${\bf Y}/\sqrt{X^2/\nu}$ is multivariate $t$, where
$X^2$ is a chi-squared random variate with $\nu$ degrees of freedom.
\section{Prototype software}
The covariance matrix computation forms part of a program that calls Numerical Algorithms Group (NAG) library routines to maximise a likelihood function.
Here a function to evaluate the log-likelihood is available, and it is assumed that iteration has reached convergence.
Three subroutines were written:
\begin{enumerate}
\item Routine getcurv: evaluates the Hessian matrix using the algorithms described here.
\item Routine newgetcov: organizes computation of the covariance matrix from the Hessian,
calling getcurv and inverting the Hessian via diagonalizing the matrix and applying eigenvalue polish
if required.
\item Routine getcov2: computes the mean and variance of posterior distribution of ${\boldsymbol \theta}$.
The maximum-likelihood parameter estimates and the Hessian and covariance matrix must have been already computed.
\end{enumerate}

\section{Examples: Analyses of burns data and basketball data}
To evaluate the covariance matrix for the maximum-likelihood estimator, one must have the matrix computed analytically for comparison.
It is straightforward (but tedious) to obtain second derivatives for a simple model. This has been done for the two examples below, and the derivative matrix is not given
here.
In each case, the Hessian, the matrix of second derivatives,  was computed analytically and of course inverted numerically.
The burns data example is an analysis with 15 parameters, maybe typical of much statistical work, and the basketball example has 322 parameters,
giving a large covariance matrix. The interest here is not in the results of the analyses, but simply in the accuracy of the computations.

In general, model parameters that can only take positive values have been transformed to logarithms where necessary, so that the transformed variables are defined on the whole real line.

In comparing covariance matrices with the analytic solution ${\bf C}$, several metrics were used. The Frobenius distance 
was taken as $F=\sum_{i=1}^n \sum_{j=1}^n |C_{ij}-V_{ij}|/n^2$, with the analogous distance per element for correlation matrices.
Also, the quantity $G=100\sum_{i=1}^n |\frac{\sqrt{V_{ii}}-\sqrt{C_{ii}}}{n\sqrt{C_{ii}}}|$ is the average percentage difference between analytic and computed standard errors.
This is the quantity of most interest to the majority of practitioners, who simply want accurate standard errors on fitted model parameters.
It is only occasionally that the full matrix ${\bf V}$ is required, \eg for computing the standard error of a derived quantity of interest that is a function
of ${\boldsymbol \theta}$. In this case, the error may be computed approximately using the delta-method, or more accurately by simulating data using $\hat{\boldsymbol\theta}$ and ${\bf V}$ and assuming multivariate normality.
\subsection{Algorithms used}
Four algorithms were evaluated:
\begin{enumerate}
\item The computations are compared with the results of using the  NAG routine E04XAF. This is a recoding of the FDCALC/FDCORE routines of Gill (1983).
This was not designed to produce very accurate estimates of the Hessian, but rather to give a quick computation of it for use in function optimization. However, currently it is the only such offering in the NAG library.
\item The standard method is the computation of diagonal elements of the Hessian using Ridders' method, and the off-diagonal elements applying Richardon extrapolation once.
\item The `standard plus polish' method is as above, but with the eigenvalues of the Hessian re-evaluated. This should ensure that the matrix is always positive-definite.
\item the `quick off-diagonal' method is as per the standard method, but off-diagonal elements are not improved by Richardson extrapolation. This almost halves the total number of function evaluations,
but sacrifices accuracy.
\end{enumerate}
\subsection{Burns data}
The data and the analysis are described in Baker {\em et al} (2009), and are 768 cases of burns, the outcome variable being whether recovery took 1-13 days (group 1),
14-21 days (group 2), or over 21 days (group 3).
Here we consider 13 covariates such as burn site, gender, race, type of burn, and percentage of total body surface burnt.
The model fitted is the proportional odds (PO) model; see \eg Harrell (2001).
Writing $z={\boldsymbol \beta}^T{\bf x}$ for the effect of covariates ${\bf x}$, the group probabilities are
\[P_{< 3}=\frac{1}{1+\exp(-z-\gamma_1-\gamma_2)},\]
\[P_{<2}\equiv P_1=\frac{1}{1+\exp(-z-\gamma_1)}.\]
The two intercept parameters $\gamma_1, \gamma_2$ give a model with 15 parameters.

Table \ref{tab1} shows measures of the accuracy of the various methods used, and of the amount of computation required.
\begin{table}[h]
\begin{tabular}{|l|c|c|c|c|c|} \hline
Method & $\Delta H$ & Corr. dist. & G & Time & Evals \\ \hline
E04XAF & 0.572 & $1.6 \times 10^{-5}$ & 0.00206 & 0.0312&466 \\ \hline
Standard & $3.5 \times 10^{-4}$ & $1.08 \times 10^{-8}$& $9.5 \times 10^{-7}$ & 0.078&1048\\ \hline
Standard plus polish&$3.5 \times 10^{-4}$ & $9.8 \times 10^{-9}$ & $5.6 \times 10^{-7}$ & 0.0936 & 1246 \\ \hline
Quick off-diag. & 3.14& $8.79 \times 10^{-5}$ & 0.0118 & 0.0624&628 \\ \hline
\end{tabular}
\caption{\label{tab1} Frobenius distance between Hessians, ditto for correlation matrices, percentage error on standard errors $G$, computing time in seconds, and number of function evaluations
needed for various methods of computing the covariance matrix, for the burns data.}
\end{table}
\subsection{Basketball data}
The data are 3195 games of basketball played in 2017. In ongoing work, a simple model has been fitted, where in a match the mean score from team 1 would be $t_1=\delta k^{-1}\alpha_1/\alpha_2$,
and the mean team 2 score is $t_2=k^{-1}\alpha_2/\alpha_1$. Here $k$ is a constant, and $\alpha_1, \alpha_2$ are the team strengths. If the match is home/away, $\delta$ is added as the home team advantage.
Out of all the 319 team strengths, one has to be fixed and was set to unity. The model therefore has $319-1+4=322$ parameters.

The model is that the scores $X_1, X_2$ are normally distributed, with the sum $X_1+X_2$ having variance $\sigma_s^2$, the difference $X_1-X_2$ having variance $\sigma_d^2$,
and sum and difference being independent. From one match, the likelihood contribution is
\[{\cal L}=\frac{1}{\pi\sigma_s\sigma_d}\exp(-1/2\{(X_1-X_2-(t_1-t_2))^2/\sigma_d^2+(X_1+X_2-t_1-t_2)^2/\sigma_s^2\}).\]
This of course is a special case of a bivariate normal distribution for the scores.

Table \ref{tab2} shows the results.
Note that E04XAF returned a Hessian with a negative eigenvalue, which was then `polished' to make it positive, so that the matrix could be inverted
to give the results for the computed covariance matrix shown.

\begin{table}[h]
\begin{tabular}{|l|c|c|c|c|c|} \hline
Method & $\Delta H$ & Corr. dist. & G & Time & Evals \\ \hline
E04XAF & 0.446146& 0.0356 & 2.73& 15.26&157622 \\ \hline
Standard & $3.16 \times 10^{-7}$&$3.39 \times 10^{-8}$ &$2.90 \times 10^{-6}$&39.81&417018 \\ \hline
Standard plus polish&$1.5 \times 10^{-7}$&$5.56 \times 10^{-9}$&$1.32 \times 10^{-7}$&39.65&420528 \\ \hline
Quick off-diag. & $6.02 \times 10^{-4}$&$2.35 \times 10^{-3}$&0.2041 &19.71&210294\\ \hline
\end{tabular}
\caption{\label{tab2} Frobenius distance between Hessians, ditto for correlation matrices, percentage error on standard errors $G$, computing time in seconds, and number of function evaluations
needed for various methods of computing the covariance matrix, for the basketball data.}
\end{table}

\subsection{Bayesian computations}
100000 simulations were done to find the mean and variance of the posterior distribution for the burns dataset, using the methodology described earlier.

Besides the parameters $\gamma_1, \gamma_2$, the 13 model parameters ${\boldsymbol \beta}$ comprise demographic variates such as age, gender and skin colour,
percentage of body surface area burned, position on the body, and cause of burn. The body position is relative to burns on limbs, and burn cause is relative to scalds.
A negative coefficient indicates that recovery is slower. Level of significance is shown as an asterisk for $p< 0.05$, while two asterisks indicates $p<0.01$ and three indicates $p<0.001$.
Thus burns on the face heal significantly faster, women heal more slowly, and electrical and contact burns heal more slowly.
Age matters, with patients healing faster up to about age 25, then increasingly slowly with age.

Results of fitting the model are shown in table \ref{tab3}, and compared with the maximum-likelihood estimates.
It can be seen that the difference between the MLE and the expected-likelihood estimates is usually small, and that the error usually but not invariably increases slightly.

The computational error on the means was estimated as roughly 0.1\% of their estimated value.
The progress of the adaptive iteration was that it was done in 10 batches of 10000 simulations. For the first batch, $\nu$ was 6,
for the second it was 8, and subsequently it was again 6. This took 200000 function evaluations, compared to 628 for computation of the Hessian.
Hence the approximate computation of the posterior variance takes roughly 318 times longer than the computation of the Hessian and its inverse.
This is a lot more computation, but there seems no more efficient way of computing multivariate integrals (where $n$ exceeds 3 or 4) than `clever' Monte-Carlo methods.

\begin{table}[h]
\begin{tabular}{|l|c|c|c|c|c|} \hline
Parameter & sig.&$\hat\theta$& $\tilde\theta$& $\hat\sigma$ & $\tilde\sigma$  \\ \hline
$\gamma_1$ &-&1.113&1.125&.267& .271 \\ \hline
$\gamma_2-\gamma_1$&-&0.962&0.970&.0915&.0927 \\ \hline
Extremity&***&0.905&0.919&.226&.230 \\ \hline
Torso&*&0.643&0.552&.254&.258 \\ \hline
Face&***&1.944&2.023&.514&.522 \\ \hline
Gender&***&-0.600&-0.605&.198&.201 \\ \hline
Skin colour&&0.542&0.628&.677&.690 \\ \hline
Age &&-0.0069&-0.0070&.0042&.0042 \\ \hline
$(\text{Age}-35)^2$&***&-0.00055&-0.00055&.000163&.000165 \\ \hline
\% TBSA&&-0.0123&-0.0124&.0088&.0090 \\ \hline
Flame&*&-0.434&-0.440&.221&.224 \\ \hline
Chemical&&-0.818&-0.814&.502&.512 \\ \hline
Flash&&-0.273&-0.260&.441&.448 \\ \hline
Electrical&***&-3.277&-3.533&1.163&1.120 \\ \hline
Contact&***&-1.435&-1.442&.513&.523 \\ \hline
\end{tabular}
\caption{\label{tab3} Parameter values and standard errors for the burns dataset. The circumflex denotes maximum-likelihood estimates with standard errors from the Hessian,
the tilde denotes the corresponding quantities from the posterior distribution.}
\end{table}
\section{Timing and sample size}
It is useful to understand how  the  computation time to achieve a given accuracy for the posterior mean and variance depends on sample size $N$.
The likelihood ${\cal L}=\exp(\ell)$ can be expanded in a Taylor series as
\[{\cal L} =\exp\{\ell_0-\frac{{\boldsymbol \phi}^T{\bf V}^{-1}{\boldsymbol \phi}}{2}+A+\cdots\},\]
where ${\boldsymbol \phi}={\boldsymbol \theta}-\hat{\boldsymbol \theta}$, and
\[A=(1/6)\sum_{i=1}^n\sum_{j=1}^n\sum_{k=1}^n (\partial^3\ell/\partial\phi_i\partial\phi_j\partial\phi_k)\phi_i\phi_j\phi_k,\]

Derivatives are evaluated at ${\boldsymbol \phi}={\bf 0}$ and $\ell_0=\ell({\bf 0})$.
Expanding out the exponential, we obtain the further approximation
\[{\cal L} \simeq\exp\{\ell_0-\frac{{\boldsymbol \phi}^T{\bf V}^{-1}{\boldsymbol \phi}}{2}\}(1+A).\]
Hence we have that
\[\tilde{\boldsymbol \theta}-\hat{\boldsymbol \theta} \simeq \frac{\int\exp\{-\frac{{\boldsymbol \phi}^T{\bf V}^{-1}{\boldsymbol \phi}}{2}\}(1+A){\boldsymbol\phi}\dee {\boldsymbol \phi}}{\int\exp\{-\frac{{\boldsymbol \phi}^T{\bf V}^{-1}{\boldsymbol \phi}}{2}\}(1+A)\dee {\boldsymbol \phi}}.\]
Here $\dee {\boldsymbol \phi}$ denotes $\prod_{i=1}^n\dee \phi_i$.
Using the symmetry of the normal distribution, this simplifies to
\[\tilde{\boldsymbol \theta}-\hat{\boldsymbol \theta} \simeq \frac{\int\exp\{-\frac{{\boldsymbol \phi}^T{\bf V}^{-1}{\boldsymbol \phi}}{2}\}A{\boldsymbol\phi}\dee {\boldsymbol \phi}}{\int\exp\{-\frac{{\boldsymbol \phi}^T{\bf V}^{-1}{\boldsymbol \phi}}{2}\}\dee {\boldsymbol \phi}}.\]
For the $m$th component, the numerator is a sum over terms $(\phi_i\phi_j)(\phi_k\phi_m)$. 
The likelihood derivative $\partial^3\ell/\partial\phi_i\partial\phi_j\partial\phi_k$ is $\propto N$,
and the integral over the product of 4 $\phi$'s is a product of two variance components. Since the variance $\propto 1/N$, $\tilde{\boldsymbol \theta}-\hat{\boldsymbol \theta} \propto 1/N$.
Hence the computational variance on this is $\propto 1/N^2$. Computing time per function evaluation $\propto N$, so computing time for a given accuracy $\propto 1/N$.
Thus, for very large sample sizes, the posterior mean is almost identical to the maximum-likelihood estimate, so very little computation is needed.
In the limit, none is required, as we just use the MLE estimates.

Similar calculations can be done for studying computing time for the variance $\tilde {\bf V}$. Here we also need
\[B=(1/24)\sum_{i=1}^n\sum_{j=1}^n\sum_{k=1}^n\sum_{l=1}^n(\partial^4\ell/\partial\phi_i\partial\phi_j\partial\phi_k\partial\phi_l)\phi_i\phi_j\phi_k\phi_l.\]
Then
\[\tilde{\bf V} \simeq \frac{\int\exp\{-\frac{{\boldsymbol \phi}^T{\bf V}^{-1}{\boldsymbol \phi}}{2}\}(1+A+B){\boldsymbol\phi}{\boldsymbol\phi}^T\dee {\boldsymbol \phi}}{\int\exp\{-\frac{{\boldsymbol \phi}^T{\bf V}^{-1}{\boldsymbol \phi}}{2}\}(1+A+B)\dee {\boldsymbol \phi}}-\Delta{\boldsymbol \theta}\Delta{\boldsymbol \theta}^T,\]
where $\Delta{\boldsymbol \theta}=\tilde{\boldsymbol \theta}-\hat{\boldsymbol \theta}$.
Using symmetry, the terms in $A$ vanish, and expanding out the denominator to first order we obtain
\[\tilde{\bf V}-{\bf V} \simeq \frac{\int \exp\{-\frac{{\boldsymbol \phi}^T{\bf V}^{-1}{\boldsymbol \phi}}{2}\}(B-B_a){\boldsymbol \phi}{\boldsymbol \phi}^T\dee {\boldsymbol \phi}}{
\int \exp\{-\frac{{\boldsymbol \phi}^T{\bf V}^{-1}{\boldsymbol \phi}}{2}\}\dee {\boldsymbol \phi}},\]
where 
\[B_a=\frac{\int\exp\{-\frac{{\boldsymbol \phi}^T{\bf V}^{-1}{\boldsymbol \phi}}{2}\}B\dee {\boldsymbol \phi}}
{\int\exp\{-\frac{{\boldsymbol \phi}^T{\bf V}^{-1}{\boldsymbol \phi}}{2}\}\dee {\boldsymbol \phi}}.\]
Hence the integral over $B{\boldsymbol \phi}{\boldsymbol \phi}^T$ and the $\Delta{\boldsymbol \theta}\Delta{\boldsymbol \theta}^T$ term is proportional to $N^{-2}$, the computational variance is $\propto N^{-4}$ and computation of $\tilde{\bf V}$ takes time $O(N^{-3})$.
\section{Conclusions}
Algorithms have been presented and evaluated for computing the covariance matrix on fitted model parameters, both from the inverse of the Hessian computed at the
maximum likelihood estimate $\hat{\boldsymbol \theta}$ and for the covariance matrix of the posterior distribution of $\boldsymbol \theta$.
Standard methods of numerical analysis are used, but their combination to deliver an accurate covariance matrix is original.

In Bayesian work, any prior pdf can multiply the likelihood, to give maximum a posteriori (MAP) estimates, which can then be treated as maximum likelihood estimates for the computations.
Hence this is a general method for computing the mean and variance of Bayesian estimates, which works whenever the MAP estimator exists.

The posterior distribution computations of course take much longer than the inverse Hessian computation, but this should not be a problem in general, because for small samples function evaluation is quick, and for large samples
the deviation from the maximum likelihood estimates will be small. The crude calculation in the previous section suggests that computation time for a given accuracy for the posterior mean should decrease inversely with sample size for large sample sizes, and for computation of the standard error on a model parameter, 
time needed decreases as the cube of the sample size.

\end{document}